\documentclass[12pt,twoside]{article}
\usepackage{xrb2007}
\usepackage{epsf}
\usepackage{epsfig}

\begin{document}

\session{Future Missions and Surveys}

\shortauthor{Lewis, Russell, Fender and Roche}
\shorttitle{Monitoring LMXBs with the Faulkes Telescopes}

\title{Monitoring LMXBs with the Faulkes Telescopes}
\author{Fraser Lewis and Paul Roche}
\affil{LCOGTN, Cardiff University, 5, The Parade, Cardiff, CF24 3AA}
\affil{The Open University, Walton Hall, Milton Keynes, MK7 6AA}
\author{David M. Russell and Rob P. Fender}
\affil{University of Amsterdam, Kruislaan 404, 1098 SM Amsterdam}
\affil{University of Southampton, Highfield, Southampton, SO17 1BJ}

\begin{abstract}
We have been undertaking a monitoring project of 13 low-mass X-ray binaries (LMXBs) using FT North since early 2006. The introduction of FT South has allowed us to extend this monitoring to include 15 southern hemisphere LMXBs (see Figure 1). With new instrumentation, we also intend to expand this monitoring to include both infrared wavelengths and spectroscopy.
\end{abstract}
\section{Introduction}
The Faulkes Telescope Project is an educational and research arm of the Las Cumbres Observatory Global Telescope Network (LCOGTN). It has two 2-metre robotic telescopes, located at Haleakala on Maui (FT North) and Siding Spring in Australia (FT South). It is planned for these telescopes to be complemented by a research network of eighteen 1-metre telescopes located at 6 sites, along with an educational network of twenty-eight 0.4-metre telescopes, located at 7 sites. This will provide 24 hour coverage of both northern and southern hemispheres.
We monitor the brighter systems once per week in V, R and i' bands, and the dimmer systems in the  i' band. During outbursts or after ATels, we are able to amend this accordingly to increase the observing cadence. Sample results are shown in sections 3 and 4.

\section{Aims}
1. To identify transient outbursts in LMXBs. LMXBs may brighten in the optical/near-infrared (OIR) for up to a month before X-ray detection. The behaviour of the optical rise is poorly understood, especially for black hole X-ray binaries. Catching outbursts from quiescence will allow us to examine this behaviour and alert the astronomical community to initiate multi-wavelength follow-up observations.\\
2. To study the variability in quiescence. Recent results have suggested that many processes may contribute to the quiescent optical emission, including emission from the jets in black hole systems, e.g. \citep{ru06}. By monitoring the long-term variability of quiescent LMXBs, we aim to provide constraints on the emission processes and the mass functions.\\
\begin{figure}[!ht]
\plotfiddle{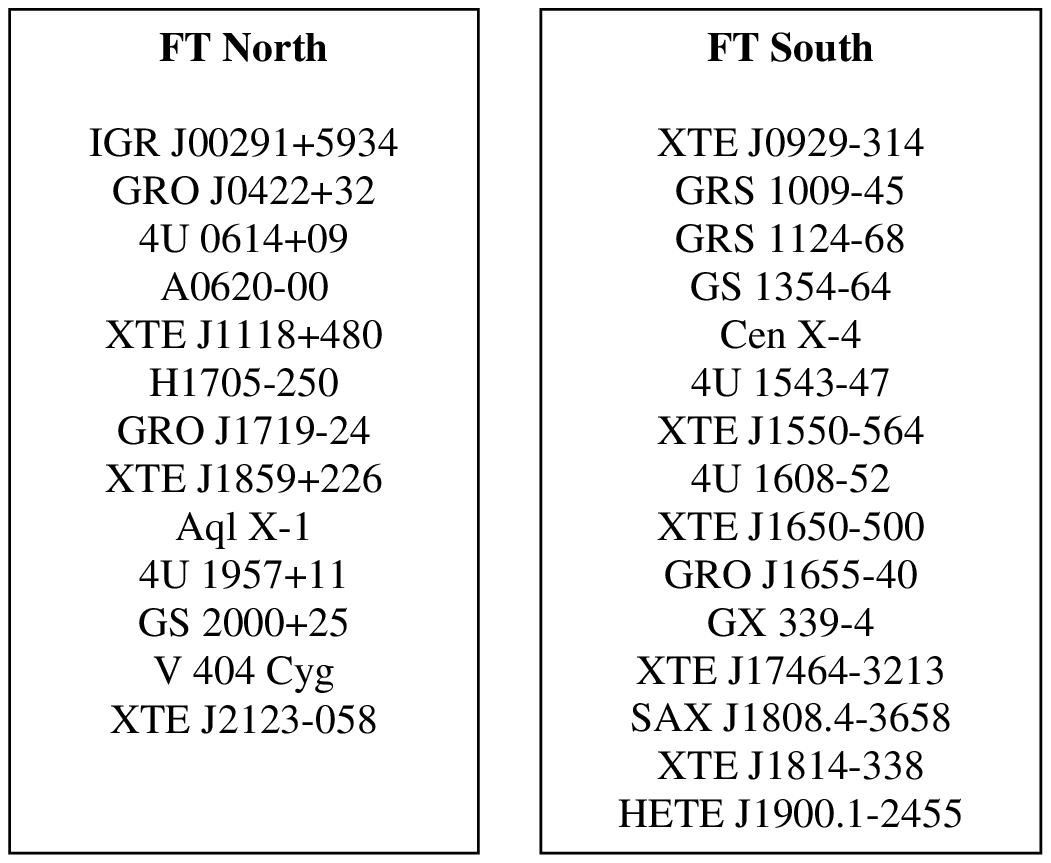}{2in}{0}{60}{60}{-250}{-300}
\caption{List of LMXB Targets}
\end{figure}
\section{GX339-4}
A target for FT South is the transient black hole binary GX339-4, which went into outburst in early 2007, followed by a steady decline in the following months. The outburst was detected at X-ray \citep{kr06}, OIR \citep{bb07} and radio \citep{co07} wavelengths. Our observations show that the source continues to decline in both V and i' bands (see Figure 2).

The de-reddened optical spectral energy distribution (SED) during this outburst decay shows a blue spectrum with an excess in the i' band. This excess is weaker in a VLT SED from a low luminosity state in 2005, (Russell and Fender 2008, MNRAS, submitted) as seen in the diagram.

Similar results were obtained from GX339-4 during the decline of the 2002 outburst \citep{ho05}, where the excess was attributed to the jet. The results indicate that the OIR contribution from the jet decreases at low luminosities in the hard state, implying that in quiescence the jet will only be detected at lower frequencies, in the mid-infrared. This is consistent with multi-wavelength quiescent observations of a number of LMXBs \citep{ga07}.

\begin{figure}[!ht]
\plotfiddle{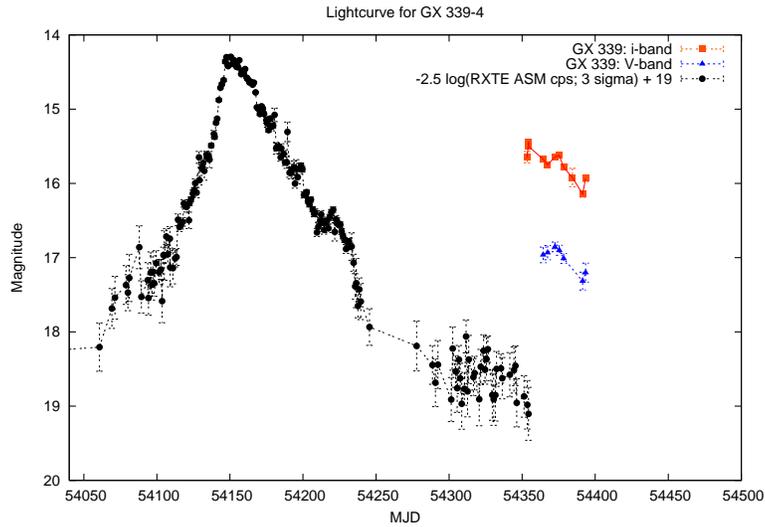}{2.5in}{270}{40}{40}{-200}{220}
\caption{Long-term light curve of GX 339-4}
\end{figure}

\section{Aquila X-1}
Aquila X-1, a transient atoll-type LMXB, went into outburst on May 20, 2007  as observed by INTEGRAL \citep{ro07} and also in the OIR \citep{mb07}. Since this time, we have been intensively monitoring this target and have observed a more recent burst of activity in V, R and i' bands \citep{ma07}.

Our observations show that in this source at least, the optical and X-ray detections of a new outburst are within a few days of each other (see Figure 3).
\section{Conclusions}
We will continue to monitor these sources in the coming years, complimenting efforts by other teams (e.g. Buxton and Bailyn 2007) and provide long-term light curves that may reveal interesting accretion activity. Already, we detect intrinsic variability in A0620-00, XTE J1118+480, GRO J0422+32 and 4U1957+11.

\begin{figure}[!ht]
\plotfiddle{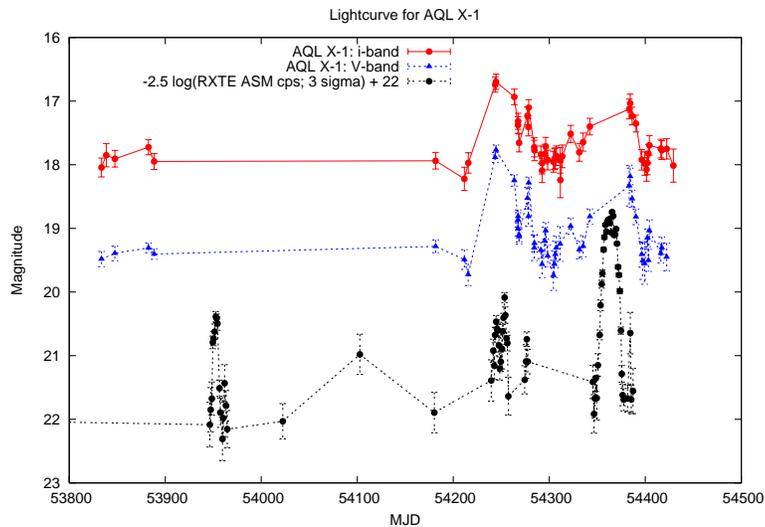}{2.5in}{270}{40}{40}{-200}{220}
\caption{Long-term light curve of Aql X-1}
\end{figure}
\acknowledgements FL would like to acknowledge support from the Dill Faulkes\\ Educational Trust.

\end{document}